\documentstyle[twoside]{article}

\catcode`\@=11
\long\def\@makefntext#1{
\protect\noindent \hbox to 3.2pt {\hskip-.9pt
$^{{\eightrm\@thefnmark}}$\hfil}#1\hfill}               

\def\@makefnmark{\hbox to 0pt{$^{\@thefnmark}$\hss}}    

\def\ps@myheadings{\let\@mkboth\@gobbletwo
\def\@oddhead{\hbox{}
\rightmark\hfil\eightrm\thepage}
\def\@oddfoot{}\def\@evenhead{\eightrm\thepage\hfil
\leftmark\hbox{}}\def\@evenfoot{}
\def\sectionmark##1{}\def\subsectionmark##1{}}

\oddsidemargin=\evensidemargin
\newcounter{sectionc}\newcounter{subsectionc}\newcounter{subsubsectionc}
\renewcommand{\section}[1] {\vspace{12pt}\addtocounter{sectionc}{1}
\setcounter{subsectionc}{0}\setcounter{subsubsectionc}{0}\noindent
        {\tenbf\thesectionc. #1}\par\vspace{5pt}}
\renewcommand{\subsection}[1] {\vspace{12pt}\addtocounter{subsectionc}{1}
      \setcounter{subsubsectionc}{0}\noindent
      {\bf\thesectionc.\thesubsectionc.{\kern1pt \bfit #1}}\par\vspace{5pt}}
\renewcommand{\subsubsection}[1]
      {\vspace{12pt}\addtocounter{subsubsectionc}{1}
      \noindent{\tenrm\thesectionc.\thesubsectionc.\thesubsubsectionc.
      {\kern1pt \tenit #1}}\par\vspace{5pt}}
\newcommand{\nonumsection}[1] {\vspace{12pt}\noindent{\tenbf #1}
        \par\vspace{5pt}}

\newcounter{appendixc}
\newcounter{subappendixc}[appendixc]
\newcounter{subsubappendixc}[subappendixc]
\renewcommand{\thesubappendixc}{\Alph{appendixc}.\arabic{subappendixc}}
\renewcommand{\thesubsubappendixc}
        {\Alph{appendixc}.\arabic{subappendixc}.\arabic{subsubappendixc}}

\renewcommand{\appendix}[1] {\vspace{12pt}
        \refstepcounter{appendixc}
        \setcounter{figure}{0}
        \setcounter{table}{0}
        \setcounter{lemma}{0}
        \setcounter{theorem}{0}
        \setcounter{corollary}{0}
        \setcounter{definition}{0}
        \setcounter{equation}{0}
        \renewcommand{\thefigure}{\Alph{appendixc}.\arabic{figure}}
        \renewcommand{\thetable}{\Alph{appendixc}.\arabic{table}}
        \renewcommand{\theappendixc}{\Alph{appendixc}}
        \renewcommand{\thelemma}{\Alph{appendixc}.\arabic{lemma}}
        \renewcommand{\thetheorem}{\Alph{appendixc}.\arabic{theorem}}
        \renewcommand{\thedefinition}{\Alph{appendixc}.\arabic{definition}}
        \renewcommand{\thecorollary}{\Alph{appendixc}.\arabic{corollary}}
        \renewcommand{\theequation}{\Alph{appendixc}.\arabic{equation}}
        \noindent{\tenbf Appendix \theappendixc #1}\par\vspace{5pt}}
\newcommand{\subappendix}[1] {\vspace{12pt}
        \refstepcounter{subappendixc}
        \noindent{\bf Appendix \thesubappendixc. {\kern1pt \bfit #1}}
        \par\vspace{5pt}}
\newcommand{\subsubappendix}[1] {\vspace{12pt}
        \refstepcounter{subsubappendixc}
        \noindent{\rm Appendix \thesubsubappendixc. {\kern1pt \tenit #1}}
        \par\vspace{5pt}}

\newcommand{\smalllineskip}{\baselineskip=10pt}

\def\eightcirc{
\begin{picture}(0,0)
\put(4.4,1.8){\circle{6.5}}
\end{picture}}
\def\eightcopyright{\eightcirc\kern2.7pt\hbox{\eightrm c}}

\def\abstracts#1#2#3{{
        \centering{\begin{minipage}{4.5in}\baselineskip=10pt\footnotesize
        \parindent=0pt #1\par
        \parindent=15pt #2\par
        \parindent=15pt #3
        \end{minipage}}\par}}

\renewenvironment{thebibliography}[1]
        {\frenchspacing
         \ninerm\baselineskip=11pt
         \begin{list}{\arabic{enumi}.}
        {\usecounter{enumi}\setlength{\parsep}{0pt}
         \setlength{\leftmargin 12.7pt}{\rightmargin 0pt} 
         \setlength{\itemsep}{0pt} \settowidth
        {\labelwidth}{#1.}\sloppy}}{\end{list}}

\newcounter{itemlistc}
\newcounter{romanlistc}
\newcounter{alphlistc}
\newcounter{arabiclistc}

\newcommand{\fcaption}[1]{
        \refstepcounter{figure}
        \setbox\@tempboxa = \hbox{\footnotesize Fig.~\thefigure. #1}
        \ifdim \wd\@tempboxa > 5in
           {\begin{center}
        \parbox{5in}{\footnotesize\smalllineskip Fig.~\thefigure. #1}
            \end{center}}
        \else
             {\begin{center}
             {\footnotesize Fig.~\thefigure. #1}
              \end{center}}
        \fi}

\newcommand{\tcaption}[1]{
        \refstepcounter{table}
        \setbox\@tempboxa = \hbox{\footnotesize Table~\thetable. #1}
        \ifdim \wd\@tempboxa > 5in
           {\begin{center}
        \parbox{5in}{\footnotesize\smalllineskip Table~\thetable. #1}
            \end{center}}
        \else
             {\begin{center}
             {\footnotesize Table~\thetable. #1}
              \end{center}}
        \fi}

\def\@citex[#1]#2{\if@filesw\immediate\write\@auxout
        {\string\citation{#2}}\fi
\def\@citea{}\@cite{\@for\@citeb:=#2\do
        {\@citea\def\@citea{,}\@ifundefined
        {b@\@citeb}{{\bf ?}\@warning
        {Citation `\@citeb' on page \thepage \space undefined}}
        {\csname b@\@citeb\endcsname}}}{#1}}

\newif\if@cghi
\def\cite{\@cghitrue\@ifnextchar [{\@tempswatrue
        \@citex}{\@tempswafalse\@citex[]}}
\def\citelow{\@cghifalse\@ifnextchar [{\@tempswatrue
        \@citex}{\@tempswafalse\@citex[]}}
\def\@cite#1#2{{$\null^{#1}$\if@tempswa\typeout
        {IJCGA warning: optional citation argument
        ignored: `#2'} \fi}}

\def\@refcitex[#1]#2{\if@filesw\immediate\write\@auxout
        {\string\citation{#2}}\fi
\def\@citea{}\@refcite{\@for\@citeb:=#2\do
        {\@citea\def\@citea{, }\@ifundefined
        {b@\@citeb}{{\bf ?}\@warning
        {Citation `\@citeb' on page \thepage \space undefined}}
        \hbox{\csname b@\@citeb\endcsname}}}{#1}}

\def\@refcite#1#2{{#1\if@tempswa\typeout
        {IJCGA warning: optional citation argument
        ignored: `#2'} \fi}}

\def\refcite{\@ifnextchar[{\@tempswatrue
        \@refcitex}{\@tempswafalse\@refcitex[]}}

\def\pmb#1{\setbox0=\hbox{#1}
        \kern-.025em\copy0\kern-\wd0
        \kern.05em\copy0\kern-\wd0
        \kern-.025em\raise.0433em\box0}

\def\fnt#1#2{\footnotetext{\kern-.3em
        {$^{\mbox{\scriptsize #1}}$}{#2}}}

\def\fpage#1{\begingroup
\voffset=.3in
\thispagestyle{empty}\begin{table}[b]\centerline{\footnotesize #1}
        \end{table}\endgroup}

\def\runninghead#1#2{\pagestyle{myheadings}
\markboth{{\protect\footnotesize\it{\quad #1}}\hfill}
{\hfill{\protect\footnotesize\it{#2\quad}}}}
\font\tenrm=cmr10
\font\tenit=cmti10
\font\tenbf=cmbx10
\font\bfit=cmbxti10 at 10pt
\font\ninerm=cmr9

\font\eightrm=cmr8

\def\qed{\hbox{${\vcenter{\vbox{                      
   \hrule height 0.4pt\hbox{\vrule width 0.4pt height 6pt
   \kern5pt\vrule width 0.4pt}\hrule height 0.4pt}}}$}}

\begin{document}

\runninghead{R. Amorim et al}
{Field-Antifield Formalism$\ldots$}

\thispagestyle{empty}\setcounter{page}{263}
\vspace*{0.88truein}
\fpage{263}

\centerline{\bf
FIELD-ANTIFIELD FORMALISM IN A NON-ABELIAN THEORY}
\vspace*{0.035truein}
\centerline{\bf WITH ONE AND TWO FORM GAUGE FIELDS}
\vspace*{0.035truein}
\centerline{\bf COUPLED IN A TOPOLOGICAL WAY}

\vspace*{0.035truein}

\vspace*{0.37truein}
\centerline{\footnotesize R. Amorim
and J. Barcelos-Neto}

\centerline{\footnotesize \it
Instituto de F\'{\i}sica,
Universidade Federal do Rio de Janeiro}
\baselineskip=10pt
\centerline{\footnotesize \it
RJ 21945-970 - Caixa Postal 68528 - Brasil}
\baselineskip=10pt
\centerline{\footnotesize \it
E-mail: amorim@if.ufrj.br , barcelos@if.ufrj.br}

\baselineskip 5mm

\vspace*{0.21truein}

\abstracts{ We make a systematic development of the non-Abelian
formulation of two-form gauge fields with topological coupling with the
Yang-Mills one-form connection.  An analysis of the gauge structure,
reducibility conditions and physical degrees of freedom is presented.
We employ the Batalin-Vilkovisky formalism  to quantize the resulting
theory.}{}{}

\bigskip

$$$$

\section{\bf Introduction}

\bigskip
There has been an increasing interest for gauge theories with rank
higher than one, specially for the case of rank two
\cite{Kalb}-\cite{Barc2}. This is so, first because these theories
have an interesting structure of constraints, related to its
reducibility condition, where quantization deserves some additional
care comparing with the usual gauge theories of rank one \cite{Kaul}.
We mention that antisymmetric tensor fields also appear as one of the
massless solutions of string theories, in company with photons,
gravitons etc. \cite{Ma}. Another interesting aspect of these
theories is to provide a possible mechanism of mass generation for
gauge fields when vector and tensor fields are coupled in a
topological way \cite{Cremmer,Amorim}.

\medskip
The general idea of this mechanism resides in the following: tensor
gauge fields are antisymmetric quantities and consequently in $D=4$
they exhibit six degrees of freedom. By virtue of the massless
condition, the number of degrees of freedom goes down to four. Since
the gauge parameter is a vector quantity, this number would be zero
if all of its components were independent. This is nonetheless the
case because the system is reducible (which means that the gauge
transformations are not all independent) and we mention that the
final number of physical degrees of freedom is just one. It is
precisely this degree of freedom that can be absorbed by the vector
gauge field in the vector-tensor gauge theory in order to acquire
mass. The problem is that the reducibility condition is a property
that is naturally present just in the Abelian case. So, a directly
non-Abelian extension of this theory would not make sense because it
would exhibit no physical degrees of freedom.

\medskip
Freedmann and Townsend \cite{Freedmann} have presented an interesting
model for a non-Abelian formulation of two-form gauge fields, where
the reducibility condition is kept on-shell in a sense that the
curvature based on the non-Abelian vector field is null. This model
has been widely analyzed by many authors, mainly due to its peculiar
structure of constraints \cite{Battle}. Unfortunately, this
particular condition of zero curvature avoids the use of this model
in connection with a full Yang-Mills theory and, consequently, the
possibility of having alternative mechanisms of mass generation for
non-Abelian vector fields.

\medskip
In recent papers, it has been shown that there is a way of getting a
non-Abelian formulation for two-form gauge fields without the
condition of zero curvature \cite{Lahiri2,Hwang,Barc1,Niemi}. This is
achieved by introducing an auxiliary field, that plays the role of a
kind of St\"uckelberg field, which makes a suitable transition
between non-Abelian and Abelian limits to avoid the problem of zero
degrees of freedom. The consistency of this formulation can be
verified by using it in the mass generation of the Salam-Weinberg
theory. The obtained mass for the vector fields are effectively the
same as the one given by the spontaneous symmetry breaking
\cite{Barc2}.

\medskip
The purpose of the present paper is twofold: First we make a
systematic development of the non-Abelian formulation of two-form
gauge fields with topological coupling with the Yang-Mills one-form
connection. We also show that the model presented in Refs.
\cite{Lahiri2,Hwang,Barc1} have a reducible gauge structure . This is important because it assures that the number
of degrees of freedom of the non-Abelian theory is just the dimension
of the algebra times the number of degrees of the freedom of its
Abelian limit. Later on we use the field-antifield
formalism due to Batalin and Vilkovisky (BV) \cite{BV,H} to quantize
the resulting model.

\medskip
Our paper is organized as follows: In Sec.  2 we make a general and
brief review of the one-form gauge field in order to fix the notation
and prepare for transition for the two-form case. In Sec. 3, we
consider the Abelian two-form case and analyze the difficulties we
have in passing to the non-Abelian formulation.  In Sec. 4, we
present a revised version of the model introduced at Ref.
\cite{Lahiri2,Hwang,Barc1} to obtain the non-Abelian two-form gauge
field theory, explore its interesting gauge algebraic structure and
develop its BV quantization. We left Sec. 5 for some concluding
remarks.

\section{\bf Brief review of the the one-form gauge field
theory}
\renewcommand{\theequation}{2.\arabic{equation}}
\setcounter{equation}{0}

\bigskip
Let us start by reviewing the Yang-Mills gauge theory, here described
in terms of Lie algebra valued p-forms, in order to establish
conventions and general definitions. They will be useful in the next
sections to properly describe two-form non-Abelian gauge field
theories.  Let

\begin{equation}
A=A_\mu^aT^adx^\mu
\label{2.1}
\end{equation}

\bigskip\noindent
be the one-form connection, with values in the Lie algebra of SU(N),
whose generators are assumed to satisfy

\begin{eqnarray}
\bigl[T^a,T^b]&=&i\,f^{abc}\,T^c
\nonumber\\
{\rm Tr}\,(T^aT^b)&=&\delta^{ab}
\label{2.2}
\end{eqnarray}

\bigskip\noindent
Although (\ref{2.1}) is defined in relation to some coordinate basis
$\{dx^\mu\}$, it can be obviously referred to any other non-holonomic
basis.

\medskip
On any Lie algebra valued $p$-form $\omega$ \cite{Gilkey}

\begin{equation}
\omega=\omega^a\,T^a
\label{2.3}
\end{equation}

\bigskip\noindent
it is possible to define the exterior covariant derivative by

\begin{equation}
D\omega=d\omega-iA\wedge\omega+i(-1)^p\,\omega\wedge A
\label{2.4}
\end{equation}

\bigskip\noindent
where $d$ represents the usual exterior derivative.

\medskip
The curvature two-form

\begin{equation}
F=dA-iA\wedge A
\label{2.5}
\end{equation}

\bigskip\noindent
is such that the Biachi identities

\begin{eqnarray}
DD\,\omega&=&i\,\omega\wedge F -i\,F\wedge\omega
\nonumber\\
&\equiv&i\,[\omega,F]
\label{2.6}\\
D\,F&=&0
\label{2.7}
\end{eqnarray}

\bigskip\noindent
are satisfied for any gauge connection $A$ and algebra-valued p-form
$\omega$.  A fundamental consequence of (\ref{2.6}) is that if we
define the gauge variation of the one-form connection to be given by

\begin{equation}
\label{2.8}
\delta A=D\chi
\end{equation}

\bigskip\noindent
where $\chi$ is an arbitrary algebra valued  parameter, the curvature
two-form transforms as

\begin{eqnarray}
\delta F&=&d\,\delta A-i\,\delta A\wedge A
-i\,A\wedge\delta A
\nonumber\\
&=&D\,\delta A
\nonumber\\
&=&DD\chi
\nonumber\\
&=&i\,[\chi,F]
\label{2.9}
\end{eqnarray}

\bigskip\noindent
This implies that the action

\begin{equation}
S_0=+\,\frac{1}{2}\,{\rm Tr}\int F\wedge^\ast F
\label{2.10}
\end{equation}

\bigskip\noindent
is invariant under (\ref{2.8}), due to the cyclic property of the
trace operation. In (\ref{2.10}), the symbol $^\ast $ represents the
Hodge duality operation, so the integrand is proportional to the
oriented volume element in $M_4$, the Minkowiski space-time. To be
more precise, the  duality operation maps the p-form coordinate
basis $\{1,dx^\mu,dx^\mu\wedge dx^\nu,dx^\mu\wedge dx^\nu\wedge
dx^\rho,dx^\mu\wedge dx^\nu\wedge dx^\rho\wedge dx^\sigma\}$ into the
basis
$\{\eta,\eta^\mu,\eta^{\mu\nu},\eta^{\mu\nu\rho},\eta^{\mu\nu\rho\sigma}\}$.
In these expressions, $\eta$ is the four-form oriented volume element,
$\eta^\mu$ is a three-form, $\eta^{\mu\nu}$ is a two-form and so on.
They satisfy relations such
$dx^\mu\wedge\eta_\nu=\delta^\mu_\nu\eta\,,dx^\mu\wedge\eta_{\nu\rho}
=2\delta^\mu_{[\nu}\eta_{\rho]}$ and
$dx^\mu\wedge\eta_{\nu\rho\sigma}
=3\delta^\mu_{[\nu}\eta_{\rho\sigma]}$.  As
$F={1\over2}F_{\mu\nu}dx^\mu\wedge dx^\nu$, $^\ast
F={1\over2}F_{\mu\nu}\eta^{\mu\nu}$ and consequently $F\wedge^\ast F=-
{1\over2}F_{\mu\nu}F^{\mu\nu}\eta$.

\bigskip
An arbitrary variation of $F$, due to (\ref{2.5}), takes the form
$\delta F=D\delta A$ (see Eq. \ref{2.9}). This implies that $\delta
S=-\,{\rm Tr}\int D^\ast F\wedge\delta A$. So the Hamilton principle
applied to (\ref{2.10}) implies in the equation of motion $D^\ast
F=0$.

\medskip
As it is well known, variation (\ref{2.8}) closes in an algebra.
Actually, it is easy to verify that

\begin{equation}
[\delta_2,\delta_1]\,A
\equiv \delta_3\,A
\label{2.11}
\end{equation}

\bigskip\noindent
if one defines

\begin{equation}
\chi_3=i\,[\chi_1,\chi_2]
\label{2.12}
\end{equation}

\bigskip
The field-antifield quantization of YM theory can be easily
constructed once we consider the classical action (\ref{2.10}), the
gauge variation (\ref{2.8}) and the algebra structure contained in
(\ref{2.12}). We get the field-antifield functional generator
\cite{BV,H}

\begin{equation}
Z[J]=\int D\Phi^A\,D{\Phi}_A^\ddagger\,
\delta\,\Bigl[\Phi_A^\ddagger
-{\partial\Psi\over{\partial\Phi^A}}\Bigr]\,
\exp i\,\Bigl(S+{\rm Tr}\int J\wedge A\Bigr)
\label{2.13}
\end{equation}

\bigskip\noindent
where

\begin{equation}
\label{2.14}
S={\rm Tr}\int\Bigl(\,\frac{1}{2}\,F\wedge\ast F
-A^\ddagger\wedge Dc -i\, c^\ddagger\,c^2
+\bar c^\ddagger\wedge g\Bigr)
\end{equation}

\bigskip\noindent
and $J$ is an external three-form current. In (\ref{2.14}), $c$ is
the ghost corresponding to the parameter $\chi$, $\bar c$ is its
corresponding four-form antifield and $\bar A$ is the three-form
antifield corresponding to $A$.  $\Phi^A$ and $ \Phi_A^\ddagger$ represent
the components of all the fields and antifields appearing in $S$. The
form degrees of the antifields introduced in (\ref{2.14}) are defined
in such a way that the integrand is proportional to the oriented
volume element. They also have opposite Grassmanian parity when
compared to the corresponding fields. The pair $\bar g$ and $b$ is
necessary for the gauge-fixing procedure. This is also done with the
aid of the gauge-fixing fermion $\Psi$. According to Eq.
(\ref{2.13}), we must restrict the antifields by the condition
$\Phi_A^\ddagger={{\partial\Psi}\over{\partial \Phi^A}}$. For
non-degenerated choices, the generator functional is independent of
$\Psi$. A convenient gauge-fixing can be given by $\Psi={\rm Tr}\int
\bar cd^\ast A$. The usual Faddeev-Popov expression
for the functional generator is recovered after integrating over the
antifields and the auxiliary pairs.

\medskip
It is convenient to introduce a fundamental structure in the
field-antifield procedure which is given by the so-called
antibracket. Let $X$ and $Y$ be algebra valued forms.  The
antibracket between $X$ and $Y$ is given by

\begin{equation}
(X,Y)=\frac{\partial_rX}{\partial\Phi^A}\,
\frac{\partial_lY}{\partial{\Phi}_A^\ddagger}
-\frac{\partial_r X}{\partial{\Phi}_A^\ddagger}\,
\frac{\partial_lY}{\partial\Phi^A}
\label{2.15}
\end{equation}

\bigskip
We observe that the Witt notation of sum and integration over internal
variables are being understood. The BRST variation of any functional
is defined through

\begin{equation}
s\,X=(X,S)
\label{2.16}
\end{equation}

\bigskip\noindent
It is not difficult to verify that ${\it s}$ is  nilpotent, as a
consequence of the master equation $s\,S=0$, which is satisfied as
the theory is anomaly free. It is a mere exercise to derive the BRST
variations for all fields and antifields and verify that indeed they
act as (nilpotent) right differentials.

\section{\bf Two-form gauge field theories}
\renewcommand{\theequation}{3.\arabic{equation}}
\setcounter{equation}{0}

\bigskip
Let us start from the Abelian two-form case in order to become clear
what are the difficulties we have to pass to its non-Abelian
counterparts.  To do this, we see that in analogy to the one-form
gauge field theory, one can introduce a two-form gauge-field

\begin{equation}
B=\frac{1}{2}\,B_{\mu\nu}\,dx^\mu\wedge dx^\nu
\label{3.1}
\end{equation}

\bigskip\noindent
We note that this is not a connection. In spite of this fact we
define a geometric quantity $H$ in a similar way to the Abelian
curvature $F$, that is to say, by using the exterior derivative,

\begin{eqnarray}
H&=&dB
\nonumber\\
&\equiv&\frac{1}{6}\,H_{\mu\nu\rho}\,
dx^\mu\wedge dx^\nu\wedge dx^\rho
\label{3.2}
\end{eqnarray}

\bigskip\noindent
where

\begin{equation}
H_{\mu\nu\rho}=\partial_\mu B_{\nu\rho}
+\partial_\rho B_{\mu\nu}+\partial_\nu B_{\rho\mu}
\label{3.3}
\end{equation}

\bigskip
Concerning the gauge transformation of $B$, we assume that it has a
similar transformation to the Abelian version of that one of $A$ (see
expression (\ref{2.8})) :

\begin{equation}
\delta B=d\xi
\label{3.4}
\end{equation}

\bigskip\noindent
Here, $\xi$ is a one-form gauge parameter. We directly notice a
characteristic of the Abelian two-form formulation. Since $\xi$ is
a one-form parameter, one may rewrite it in terms of a exterior
derivative of some zero-form parameter, say

\begin{equation}
\xi=d\,\alpha
\label{3.5}
\end{equation}

\bigskip\noindent
If this is done, no gauge transformation is obtained for $B$.  This
means that the components of the gauge parameter $\xi$ are not all
independent (that is why the theory is said to be reducible).  This
is a welcome result because if this was not so, the theory written
in terms of $B$ would have zero degrees of freedom.  Actually, the
action

\begin{equation}
S_0=-\,\frac{1}{2}\int H\wedge^\ast H
\label{3.6}
\end{equation}

\bigskip\noindent
describes a reducible theory with only one degree of freedom, being
equivalent to a massless scalar field.  Actually, the numer of
degrees of freedom \cite{GPS} is $n\,(=6$: number of components of
$B$ ) $-\,n_1\,(= 2$: due to the massless condition of the field $B$
)$ -m_0\,(= 4$: number of gauge parameters) $+m_1\,(=1=$ number of
reducibility conditions), which gives $n=1$. The study of scalar
particles by means of this involved theory is natural in the context
of some string and supergravity formulations \cite{Freedmann}. An
interesting feature of such a theory is that it can effectively
generate mass for one-form gauge fields, without obstructing their
gauge invariance. In the Abelian case, this goal can be done if one
adopts the action

\begin{equation}
S_0=\int\Bigl(\,\frac{1}{2}\,F\wedge^\ast F
-\frac{1}{2}\,H\,\wedge^\ast H+m\,F\wedge B\Bigr)
\label{3.7}
\end{equation}

\bigskip\noindent
which presents a general topological coupling between the one-form
and the two-form gauge fields. Here, the number of degrees of freedom
is 3, one for $B$ and two for $A$. It is easy to verify that when one
eliminates $B$ from the two coupled equations of motion derived from
(\ref{3.7}), one verifies that $F$ satisfies a massive wave equation.
The quantization  of such a theory will not be discussed here, but it
is well known that the field $A$ presents a massive pole in its
propagator. For details, see Ref. \cite{Amorim}.

\medskip
We now pass to consider the non-Abelian case. When compared to the
one-form gauge field theory, the non-Abelian extension of the
two-form case is more subtle. The first problem arises in the gauge
transformation. Since $B$ is not a connection and we do not have a
precise definition of its geometrical nature, it appears to be
reasonable to infer that its gauge transformation is the non-Abelian
generalization of (\ref{3.4}), i.e.

\begin{equation}
\delta B=D\xi
\label{3.8}
\end{equation}

\bigskip\noindent
The point we would like to emphasize here is that the transformation
(\ref{3.8}) is not reducible. If one rewrites the one-form parameter
$\xi$ in terms of a zero-form, say $\alpha$, as $\xi=D\alpha$, one
obtains

\begin{eqnarray}
\delta B&=&\,DD\alpha
\nonumber\\
&=&i\,[\alpha,F]
\label{3.9}
\end{eqnarray}

\bigskip\noindent
where in the last step we have used the first Bianchi identity
(\ref{2.6}). We then notice that the theory can be considered
reducible just on-shell if the curvature $F$ vanishes identically
\cite{Freedmann}.

\medskip
A second problem concerns the definition of $H$. In the Abelian case,
the Maxwell curvature two-form is obtained by just taking the
exterior derivative of the connection. So, in that case, it was very
natural the obtainment of the Abelian $H$ by means of the exterior
derivative of $B$. However, the non-Abelian Yang-Mills curvature is
not the covariant derivative of the connection. Thus, we have to be
carefully in defining what is the non-Abelian $H$.

\medskip
It is usually considered the definition \cite{Lahiri2} - \cite{Barc1}

\begin{eqnarray}
H&=&DB
\nonumber\\
&=&dB-i[A,B]
\label{3.10}
\end{eqnarray}

\bigskip\noindent
which has the correct Abelian limit and also takes values in the
SU(N) algebra if $B$ does. Now, under (\ref{3.8}), for arbitrary $A$
variations, we have

\begin{eqnarray}
\delta H&=&DD\xi-i[\delta A,B]
\nonumber\\
&=&i[\xi,F]-i[\delta A,B]
\label{3.11}
\end{eqnarray}

\bigskip\noindent
So the non-Abelian generalization of (\ref{3.6}) will only be gauge
invariant if we impose that not only $F$ must vanish, but also that
$A$ must not transform.  The action \cite{Freedmann}

\begin{equation}
S_0=\,\frac{1}{2}\,{\rm Tr}\int\Bigl(F\wedge B^\ast
+{1\over2}A\wedge^\ast A\Bigr)
\label{3.12}
\end{equation}

\bigskip\noindent
presents these features.  It is invariant under $\delta B=D\xi$ when
one restricts $\delta A$ to vanish. As $F$ also vanishes as a
consequence of the equation of motion of $B$,  it describes a pure
gauge given by $A=ig^{-1}d\,g$, where $g$ represents a $SU(N)$ group
element. It is easy to show that it presents $1\times (N^2-1)$
degrees of freedom, where $N^2-1$ gives the dimension of the algebra.
Actually ({\ref{3.10}) is completely equivalent to a SU(N)
$\sigma$-model. Also, after eliminating  $A$, it has (\ref{3.6}) as
the Abelian limit. The point here is that in a theory like that, $B$
is coupled to a one-form gauge field which is a pure gauge. This fact
forbids a mechanism of mass generation for non-vanishing curvature
gauge one-form fields, that can be constructed, for instance, from
(\ref{3.7}). This is so because there the curvature two-form has a
dynamical role and can not be set to zero.

\section{\bf Coupling two-forms with non pure-gauge one-forms}
\renewcommand{\theequation}{4.\arabic{equation}}
\setcounter{equation}{0}

\bigskip
The solution of the problems quoted in the end of the last section
can be achieved by redefining the quantity $B$ by \cite{Lahiri2} -
\cite{Barc1}

\begin{equation}
\tilde B=B+D\Omega
\label{4.1}
\end{equation}

\bigskip\noindent
where the one-form $\Omega$ plays the role of a  St\"uckelberg
field. In Ref. \cite{Sorella} the same procedure is applied in the context of the BF Yang-Mills theory.

\medskip
Under the transformations

\begin{eqnarray}
\delta B&=&i[\chi,B]+D\xi+i[\eta,F]\nonumber\\
\delta \Omega&=&i[\chi,\Omega]-\xi-D\eta
\label{4.2}
\end{eqnarray}

\bigskip\noindent
we imediatelly verify that\bigskip
\begin{eqnarray}
\delta \tilde B&=&i[\chi,\tilde B]\
\nonumber\\
\delta\tilde H&=&i[\chi,\tilde H]
\label{4.3}
\end{eqnarray}

\bigskip\noindent
where $\tilde H=D\tilde B$. Transformations (\ref{4.2}) generalize
(\ref{3.8}) and is reducible, as we are going to see soon. The sector
associated with the 0-form parameter $\eta$ is an additional symmetry
related to the reducibility of the theory. Of course, we are keeping
the transformations (\ref{2.8})-(\ref{2.9}) for $A$ and $F$
respectively.

\medskip
Due to the cyclic property of the trace operation, the non-Abelian
generalization of action (\ref{3.6}) now becomes easy. It is enough
to use $\tilde H$ instead of $H$, and there is no necessity of
imposing pure gauge for the Yang-Mills sector. Furthermore, in the
Abelian limit, $\tilde H$ becomes identical to (\ref{3.10}), since
$d^2$ vanishes identically.

\medskip
The inclusion of a topological term does not break the gauge
invariance.  One can verify that the action \cite{Lahiri2,Hwang}

\begin{equation}
S_0={\rm Tr}\,\Bigl(\frac{1}{2}\,F\wedge^\ast F
-\frac{1}{2}\,\tilde H\,\wedge^\ast {\tilde H}
+m\,F\wedge \tilde B\Bigr)
\label{4.4}
\end{equation}

\bigskip\noindent
is indeed invariant under (\ref{2.8}) and (\ref{4.2}). From
(\ref{4.4}) one can extract the equations of motion, associated
respectively with the variations of $B$, $\Omega$ and $A$, as

\begin{eqnarray}
\label{4.5}
D^\ast\tilde H+mF&=&0
\nonumber\\
\left[F,^\ast{\tilde H}\right]&=&0
\nonumber\\
D(^\ast F-i[\Omega,^\ast\tilde H]-m\tilde B)
+i[B,^\ast\tilde H]&=&0
\end{eqnarray}

\bigskip\noindent
We observe that the second of the equations above gives the
integrability condition for the first of them. As $\Omega$ does not
appear dynamically in the equations of motion, it can be seen simply
as an auxiliary field, with no dynamics. This means that quantically
there will be  present  no modes associated to them. It is trivial to
verify that  equations (\ref{4.5})imply, in the Abelian limit, that

\begin{equation}
\label{4.6}
d^\ast d^{\,\ast}F+m^2\,F=0
\end{equation}

\bigskip\noindent
which means that the free theory is massive and that does not contain
the presence of the St\"uckelberg field $\Omega$.

\medskip
Let us now consider the quantization of the theory described by the
action (\ref{4.4}). First we need to derive the algebraic structure
of its gauge transformations. We can verify that the usual Yang-Mills
structure given by (\ref{2.12}) is kept, but it is also necessary to
consider that there is a mixing in the composition rules for the
other parameters. We get

\begin{eqnarray}
\xi_3&=&i\,[\chi_1,\xi_2]-i\,[\chi_2,\xi_1]\nonumber\\
\eta_3&=&i\,[\chi_1,\eta_2]-i\,[\chi_2,\eta_1]
\label{4.7}
\end{eqnarray}

\bigskip\noindent
and it is a mere exercise to show that the algebra closes on all the
fields $\phi^i$ belonging to the theory:
$\left[\delta_2,\delta_1\right]\delta\phi^i=\delta_3\phi^i$.\bigskip

\medskip
Now we observe that the gauge transformations (\ref{2.8}) and
(\ref{4.2}) are reducible. To show this, let us use a compact
notation where the fields  $\phi^i$, given here by $A,\, B$ and
$\Omega$ are represented in a column matrix

\begin{equation}\label{phi}
(\phi)=\left(\begin{array}{c}A\\B\\ \Omega
\end{array}\right)
\end{equation}

\bigskip\noindent
Consistently, the gauge parameters can also be written in the same way

\begin{equation}\label{epsilon}
(\epsilon)=\left(\begin{array}{c}\chi\\ \xi\\ \eta
\end{array}\right)\,.
\end{equation}

\bigskip\noindent
In matrix form, the gauge transformations (\ref{2.8}) and (\ref{4.2})
are then just given by

\begin{equation}\label{gauge}
\delta\,\,(\phi)=(R\,)(\epsilon)
\end{equation}

\bigskip\noindent
where the gauge generator is

\begin{equation}\label {R}(R\,)=\left(\begin{array}{clcr}D&0&0\\
i[\,\,,B]&D&i[\,\,,F]\\
i[\,\,,\Omega]&-1&-D\end{array}\right)
\end{equation}

\bigskip\noindent
In $(R)$ the lines correspond respectively to $A,\, B$ and $\Omega$
and the columns to $\chi,\,\xi$ and $\eta$.  Now we note that $\det
(R)=-D(D^2-i[\,\,,F])$, which vanishes identically due to Bianchi
identity (\ref{2.6}). This indicates that it has at least one
null vector $(Z)$, that gives  the reducibility conditions
associated with the gauge symmetries.  It is easy to verify that
the null vector is given by

\begin{equation}\label{Ztheta}
(Z)=\left(\begin{array}{c}0\\D\\ -1
\end{array}\right)
\end{equation}

Incidentally, we mention that in higher dimensions, other linearly independent null vectors can be found.

\bigskip
Following the usual rules for the field-antifield quantization
\cite{GPS,H}, we get from (\ref{2.8}), (\ref{2.11}), (\ref{4.2}),
(\ref{4.7}) and (\ref{4.13}), the functional generator (see Eq.
(\ref{2.13}))

\begin{equation}
Z[J,Q]=\int D\Phi^A\,D{\Phi}_A^\ddagger\,
\delta\Bigl[\Phi_A^\ddagger
-{\partial\Psi\over{\partial\Phi^A}}\Bigr]\,
\exp i\,\Bigl\{S+{\rm Tr}\int\left(J\wedge A
+Q\wedge B\right)\Bigr\}
\label{4.8}
\end{equation}

\bigskip\noindent
where

\begin{eqnarray}
S&=&S_0+{\rm Tr}\int\Bigl( A^\ddagger\wedge Dc
+ B^\ddagger\wedge\left(i\left[c,B\right]+Db+i\left[a,F\right]\right)+
\Omega^\ddagger\wedge
\left(i\left[c,\Omega\right]-b-D\,a\right)
\nonumber\\
&&\phantom{S_0}
-i\, c^\ddagger\,c^2+
 b^\ddagger\wedge\left(-i\{b,c\}+D\theta\right)+
 a^\ddagger\wedge\left(-i\{a,c\}-\theta\right)
\nonumber\\
&&\phantom{S_0}+\bar a^\ddagger\wedge e+\bar b^\ddagger\wedge f+
\bar c^\ddagger\wedge g +\bar\theta^\ddagger\wedge h+\Xi^\ddagger\wedge \pi
+\dots\Bigr)
\label{4.9}
\end{eqnarray}

\bigskip
Note that it was not included an  external source for $\Omega$, since
it is not dynamical, but we have introduced the two-form external
source $Q$ which couples to $B$. The action $S_0$ is given by
(\ref{4.4}). $c$ is the usual ghost corresponding to the gauge
parameter $\chi$, $a$ is the zero-form ghost corresponding to $\eta$
and $b$ is the one-form ghost corresponding to the one-form gauge
parameter $\xi$.  Since the theory has one reducibility relation, we
correspondingly have introduced the ghost-for-ghost $\theta$ .  The antifields corresponding to $B$ and $\Omega$  are
respectively two and three  forms. Those corresponding to the $c$ and
$a$ are four-forms, and that one corresponding to $b$ is a
three-form. Also trivial pairs were introduced in
order to fix the degrees of freedom related to the forms
$B$ and $\Omega$ and their reducibilities.  As usual, $\{b,c\}$ means that $b$ and $c$ are simmetrized.

\medskip
All the original fields $\phi^i$ have ghost number zero, the ghosts
have ghost number one and the ghosts-for-ghosts have ghost number
two.  The ghost number of an antifield is minus the ghost number of
the corresponding field minus one. Of course, $S$ has total ghost
number zero. Now, the dots in (\ref{4.9}) represent  terms depending
on possible higher order reducibility conditions or  higher rank
gauge structure functions \cite{GPS}. Instead of calculating them
directly from the algebraic gauge structure of the theory, they can
be determined from the classical master equation, or equivalently,
from the nilpotency of the BRST transformations over all the fields
and antifields.

\bigskip
According to (\ref{2.16}), the BRST transformations of the fields
appearing in action (\ref{4.9}) are given by

\begin{eqnarray}
s\,A&=&Dc
\nonumber\\
s\,B&=&i\,[c,B]+D\,b+i\,[a,F]
\nonumber\\
s\,\Omega&=&-\,i\,[c,\Omega]-b-D\,a
\nonumber\\
s\,c&=&-\,ic^2
\nonumber\\
s\,b&=&-\,i\,\{b,c\}+D\theta
\nonumber\\
s\,a&=&-i\{a,c\}-\theta
\label{4.10}
\end{eqnarray}

\bigskip\noindent
besides the BRST transformations that come from trivial pairs . It
is a kind of directly calculation to verify the nilpotency of the
transformations (\ref{4.10}). We have just to be a little careful
with the condition that $s$ acts as right derivative, which is a
consequence of the definition given by expression (\ref{2.16}). For
instance,

\begin{eqnarray}
s^2\,A&=&s\,Dc
\nonumber\\
&=&s\,(dc-i\,[A,c])
\nonumber\\
&=&ds\,c+i\{s\,A,c\}-i[A,s\,c]
\label{4.11}
\end{eqnarray}

\bigskip\noindent
which vanishes identically when one uses again (\ref{4.10}). The
nilpotency condition for the other fields can be verified in the same
way. It is not difficult to show that

\begin{eqnarray}
s^2\,b&=&D\left(s\theta-i[c,\theta]\right)
\nonumber\\
s^2a&=&-\left(s\theta-i[c,\theta]\right)
\label{s2}
\end{eqnarray}

\bigskip\noindent
which vanishes identically if

\begin{equation}
s\,\theta=\,i[c,\theta]
\label{stheta}
\end{equation}

\bigskip\noindent
and the $BRST$ transformations actually act as a differential over
all the fields of the theory.  The expressions above show us that we
have to add to the integrand of $S$ given in expression (\ref{4.9})
the term

\begin{equation}
i\bar\theta[c,\theta]
\label{Sdots}
\end{equation}

\bigskip\noindent
which completes the form of the BV action. At the same time, this
gives us the expression for the remaining higher order gauge
structure functions. It can be verified that this theory is anomaly
free and as a consequence the quantum action is just given by $S$,
the quantum master equation being in this way identified with the
classical one. This verification can follow the lines presented
in Ref. \cite{Sorella} and the detailed analysis will be presented elsewhere.

\medskip
Once the BV action (\ref{4.9}) has been given, we can
calculate the number of degrees of freedom. Forgetting the dimension
of the algebra, we see that $n_0=4+6$ and $n_1=1+2$, due to the
number of components of $A$ and $B$ and their massless conditions.
$m_0=1+4+1$ which is the number of gauge parameters and $m_1=1$,
since we have one reducibility condition.  This gives the number of
degrees of freedom $n_0-n_1-m_o+m_1=2$. This seems to be not the same as
the one found for the Abelian case. The point here is that the equation of motion coming from the variation of $\Omega$ is actually a constraint.
Analysis of the constraint structure of the theory shows that
the number of degrees of freedom actually is $3$\, as it should be \cite{Lahiri2}.
\medskip
To fix the gauge we also need a fermion $\Psi$. We can choose

\begin{equation}
\Psi={\rm Tr}\int\left(\bar c\,\,d^\ast\!\!\,A
+\bar b\wedge\,\,d^\ast\,\!B+\bar a\,\,d^\ast\!\, \Omega
+\bar\theta\,\,d^\ast\,\!b-\,d^\ast\,\!\bar b\wedge\Xi\right)\label{4.12}
\end{equation}

\bigskip\noindent
With this fermionic gauge-fixing functional, the antifields can be
directly calculated. Replacing these values into the
expression of $S$ we get

\begin{eqnarray}
S_{eff}&=&S_0+\int\Bigl\{\,d\,\,^\ast\!\!A\, g+\left(d\,\,^\ast\,\!B-^\ast
d\,\Xi\right)\wedge\, f+d\,\,^\ast\,\!\Omega\wedge\,e
\nonumber\\
&&\phantom{S_0}
-d\,\,^\ast\,\!\bar c\wedge Dc
-\,^\ast\!\,d\,\bar b\wedge\Bigl(i[c,B]+Db+i\left[a,F\right]\Bigr)
-d\,\,^\ast\,\!a\wedge\left(i\{a,c\}+D\theta\right)
\nonumber\\
&&\phantom{S_0}
+\,^\ast\,d\,\bar\theta\left(\{b,c\}-D\theta\right)
-\,^\ast\,d\,\bar b\wedge\pi+\,^\ast\,d\,b\wedge h
\Bigr\}
\label{4.13}
\end{eqnarray}

\bigskip\noindent 
Integrating over the fields and ghosts belonging to the $B$ sector of the theory, one obtains in the
non-interacting limit that the free vector field $A$ acquires a mass
$m$ that appears as a pole of the corresponding propagator
\cite{Amorim}.  It is important to mention that in the particular
case of Abelian groups, the field $\Omega$ decouples from the $A,\,B$
sector, which keeps the canonical form found in \cite{Kaul}.

\section{\bf Conclusion}

\bigskip
In this work we have studied the BV quantization of a non-Abelian
version of a two-form gauge field theory, where there is a
topological coupling with a nonpure gauge Yang-Mills connection. The
gauge algebraic structure of the the theory was derived, pointing
also its  reducibility  character. All of these aspects have been properly considered at the
field-antifield functional quantization.

\section{\bf Acknowledgment:} This work is supported in part by
Conselho Nacional de Desenvolvimento Cient\'{\i}fico e Tecnol\'ogico
- CNPq, Financiadora de Estudos e Projetos - FINEP, and
Funda\c{c}\~ao Universit\'aria Jos\'e Bonif\'acio - FUJB (Brazilian
Research Agencies).

\nonumsection{References}

\end{document}